\def \SAIT #1 #2 {{\em Mem.\ Soc.\ Astron.\ It.\/} {\bf #1}, #2}
\def \MESS #1 #2 {{\em The Messenger\/} {\bf #1}, #2}
\def \ASTRNACH #1 #2 {{\em Astron. Nach.\/} {\bf #1}, #2}
\def \AAP #1 #2 {{\em Astron. Astrophys.\/} {\bf #1}, #2}
\def \AAL #1 #2 {{\em Astron. Astrophys. Lett.\/} {\bf #1}, L#2}
\def \AAR #1 #2 {{\em Astron. Astrophys. Rev.\/} {\bf #1}, #2}
\def \AAS #1 #2 {{\em Astron. Astrophys. Suppl. Ser.\/} {\bf #1}, #2}
\def \AJ #1 #2 {{\em Astron. J.\/} {\bf #1}, #2}
\def \ANNREV #1 #2 {{\em Ann. Rev. Astron. Astrophys.\/} {\bf #1}, #2}
\def \APJ #1 #2 {{\em Astrophys. J.\/} {\bf #1}, #2}
\def \APJL #1 #2 {{\em Astrophys. J. Lett.\/} {\bf #1}, L#2}
\def \APJS #1 #2 {{\em Astrophys. J. Suppl.\/} {\bf #1}, #2}
\def \APSS #1 #2 {{\em Astrophys. Space Sci.\/} {\bf #1}, #2}
\def \ASR #1 #2 {{\em Adv. Space Res.\/} {\bf #1}, #2}
\def \BAIC #1 #2 {{\em Bull. Astron. Inst. Czechosl.\/} {\bf #1}, #2}
\def \JSQRT #1 #2 {{\em J. Quant. Spectrosc. Radiat. Transfer\/} {\bf #1}, #2}
\def \MN #1 #2 {{\em Mon. Not. R. Astr. Soc.\/} {\bf #1}, #2}
\def \MEM #1 #2 {{\em Mem. R. Astr. Soc.\/} {\bf #1}, #2}
\def \PLR #1 #2 {{\em Phys. Lett. Rev.\/} {\bf #1}, #2}
\def \PASJ #1 #2 {{\em Publ. Astron. Soc. Japan\/} {\bf #1}, #2}
\def \PASP #1 #2 {{\em Publ. Astr. Soc. Pacific\/} {\bf #1}, #2}
\def \NAT #1 #2 {{\em Nature\/} {\bf #1}, #2}
\def \be {\begin{equation}}
\def \ee {\end{equation}}

\documentstyle{memsait}
\input epsf.sty
\begin{opening}
\title{INNER MAGNETOSPHERIC ACCELERATORS IN HIGH MAGNETIC FIELD
 PULSARS}
\author{BING ZHANG$^1$, ALICE K. HARDING$^2$}
\institute{$^1$Astronomy \& Astrophysics Department, 
Pennsylvania State University, USA\\
$^2$Laboratory of High Energy Astrophysics, 
NASA Goddard Space Flight Center, USA}
\date{} 
\end{opening}

\def\siml{\lower4pt \hbox{$\buildrel < \over \sim$}}
\def\simg{\lower4pt \hbox{$\buildrel > \over \sim$}}

\begin{document}

\oddpagefooter{}{}{} 
\evenpagefooter{}{}{} 
\ 
\bigskip

\begin{abstract}
Photon splitting is a QED process that can potentially suppress pair
production in the inner magnetosphere of a pulsar with a super-strong
magnetic field, and hence, may quench radio emission from these
objects. While it is unknown how many splitting modes really operate
in super-critical fields, 
we derive gap parameters of a high magnetic field pulsar for both the 
vacuum-type gap and the space-charge-limited flow accelerator under 
the assumption that all three splitting modes permitted by QED operate. 
The competition between photon splitting and pair production depends
on the gap parameters, and the ``photon-splitting-dominant'' line for
both cases are derived. We discuss the implications of these results
and the possible connection of the high magnetic field pulsars with
the anomalous X-ray pulsars and soft gamma-ray repeaters which are
conjectured to be magnetars. 

\end{abstract}

\section{Introduction}
Growing evidence indicates that soft gamma-ray repeaters (SGRs) and
anomalous X-ray pulsars (AXPs) are very likely magnetars, isolated
neutron stars with surface magnetic fields of the order of $10^{14}
-10^{15}$ G (Duncan \& Thompson 1992), although consensus is not fully
achieved (see, e.g., Zhang 2001). One prominent
feature of these objects is that none of them has been firmly detected
to have pulsed radio emission, making them a distinct category from
the normal radio pulsars. Radio quiescence of magnetars has been
attributed to the function of an exotic QED process, namely photon
splitting, which, in a magnetic field in excess of roughly the
critical field strength, $B_q = 4.4
\times 10^{13}$ G, causes $\gamma$-ray photons to split into two lower 
energy photons
before they are materialized to electron-positron pairs in reaction
with the strong magnetic fields (Baring \& Harding 1998; 2001).  This
argument, however, is not certain since a complete pair-suppression
requires that both photon polarizations split. This will be realized
if all the three photon splitting modes permitted by charge-parity
invariance in QED, i.e., $\perp\rightarrow \| \|$, $\perp\rightarrow
\perp\perp$, and $\|\rightarrow \perp\|$, operate.  However, under the
weak linear vacuum dispersion limit (which may not be true in a
magnetar environment) only one splitting mode ($\perp\rightarrow
\|\|$) can fulfill the energy and momentum conservation requirements
simultaneously (Adler 1971). Recently, three high magnetic field
pulsars (HBPs) are detected to be above the ``photon splitting
deathline'' for surface emission proposed by Baring \& Harding (1998),
and one of them, PSR J1814-1744 resides very closely to an AXP, 1E
2259+586, in the $P-\dot P$ (period - spindown rate) phase space,
casting doubt on the 
effectiveness of photon splitting to suppress radio emission and the
proximity in the nature of HBPs and magnetars (Camilo et
al. 2000). Zhang \& Harding (2000), after investigating the formation
condition and the properties of the two types of inner accelerators
(space-charge-limited flow, SCLF, or vacuum gap, VG) in HBPs,
conjectured that the distinct observational difference between PSR
J1814-1744 and 1E 2259+586 is due to a geometric effect, i.e.,
different orientations of the magnetic pole with respect to the
rotational pole.  
In this interpretation and in our present paper, {\em the assumption
that all three photon splitting modes operate is still adopted}. Here we
investigate this idea in more detail by deriving explicit gap
parameters of the HBPs and by re-investigating the competition between
photon splitting and pair production for various cases.

\section{Vacuum type models: deathline for anti-parallel rotators}

If all three photon splitting modes operate, pair production is
strongly suppressed when the magnetic field is strong enough. The
condition for photon splitting dominating pair production is that the
attenuation length for photon splitting gets shorter than that of pair
production. We define a line in the $B_p-P$ (or $\dot P-P$) diagram which
satisfies this condition: the photon-splitting-dominant (PSD) line
(Eq.[\ref{PSD}]). Here $B_p$ is the surface magnetic field at the
pole for the dipolar field component.
The adoptions for the $\perp$ mode, the $\|$ mode, or the
polarization-averaged photons do not change the result significantly,
and hereafter we adopt the polarization-averaged attenuation
coefficients. A precise study of the competition between photon
splitting and pair production can only be attained through numerical
simulations (e.g. Baring \& Harding 2001). However, present
simulations have not included the pulsar inner accelerator physics,
which we intend to study here. For this purpose, we derive some
approximate analytic expressions for photon splitting.

The polarization-averaged photon splitting attenuation coefficient is
(see e.g. Harding, Baring \& Gonthier 1997)
\be
T_{\rm sp}(\epsilon)\approx {\alpha^3\over 10\pi^2}{1\over \lambda}
\left({19\over 315}\right)^2  {\cal F}(B') {B'}^6 \epsilon^5
\sin^6 \theta \simeq 0.37 {\cal F}(B') {B'}^6 \epsilon^5
\sin^6 \theta,
\label{Tsp}
\ee
where $\alpha$ is the fine-structure constant, $\lambda$ is the
Compton wavelength of the electron, and ${\cal F}(B')$ is a
strong-field modification factor (see Harding et al. 1997 for
an explicit expression). Here the photon energy $\epsilon$ is in 
units of electron rest energy, the magnetic field $B'$ is in units 
of the critical field $B_{\rm cr}$, and $\theta$ denotes the angle 
between the photon momentum and the 
magnetic field. The function ${\cal F}(B') =1$ for
$B'<<1$, but $\simeq 1.9 {B'}^{-6}$ for $B'>>1$. In the regime $B'\sim
1$ in which the photon splitting starts to suppress pair production
(which is the most interesting regime we are discussing in this paper),
one expects ${\cal F}(B') \simeq {\cal A} {B'}^{-\alpha}$, where 
$0<\alpha<6$, and ${\cal A}$ is a factor of the order of unity.
Numerical simulations give $\alpha\simeq 9/4$ (Baring \& Harding 
2001). The photon splitting attenuation length, $\lambda_{\rm sp}$, 
can be obtained by solving $\int_0^{\lambda_{\rm sp}} T_{\rm sp}(s) 
ds=1$, where $s$ is the trajectory length of the photon propagation 
which is approximately $s\simeq \rho \sin\theta$,
and $\rho$ is the curvature radius of the field line. This gives
$\lambda_{\rm sp}\simeq 1.5 {\cal A}^{-1/7} {B'}^{-15/28}
\epsilon^{-5/7} \rho^{6/7}$. For a pure dipolar geometry, one has
$\rho=9.2\times 10^7 ({\rm cm}) r_{e,6}^{1/2}P^{1/2}\xi^{-1}$,
where $r_e=10^6 ({\rm cm}) r_{e,6}$ is the emission altitude, $P$ is
the pulsar period, and $0<\xi=\Theta_s/\Theta_{\rm pc}<1$ is the
ratio of the field line magnetic colatitude at the surface to the
polar cap angle at the surface.
The splitting attenuation length is then
\be
\lambda_{\rm sp}\simeq 1.0\times 10^7 ({\rm cm}) {\cal A}^{-1/7}{B'}
^{-15/28}\epsilon^{-5/7}r_{e,6}^{3/7}P^{3/7}\xi^{-6/7}.
\label{lambdasp}
\ee
Since $1<{\cal A}<1.9$, ${\cal A}^{-1/7}$ is approximately unity, and
we will drop it out hereafter.
For the magnetic field strength considered, pair production should
occur as soon as the threshold condition is fulfilled, i.e., 
$\epsilon \sin\theta=2$. Thus the pair production attenuation length
can be approximated as
\be
\lambda_{\rm pp}=2\rho \epsilon^{-1}\simeq 1.84\times 10^8 ({\rm cm})
\epsilon^{-1} r_{e,6}^{1/2} P^{1/2} \xi^{-1}.
\label{lambdapp}
\ee
Photon splitting can cause a pulsar to ``die'' only when the inner
accelerator of the pulsar is vacuum-like, formed under the condition
of strong binding of charged particles in the neutron star surface, 
because pair formation must take place near the surface (Zhang \& Harding
2000). Therefore the so-called ``photon splitting deathline'' is only
relevant for VG accelerators. One remark is that the deathline of
Baring \& Harding (2001) is derived under several conditions: (a)
emission comes from the surface; (b) field configuration is dipole;
(c) photon energy is the ``escape energy'' for both pair
production and photon splitting. We note that although both (a) and
(b) can be still adopted, the condition (c) may be not of much physical
meaning when taking into account the inner accelerators. This is
because the typical energy of the photons produced by the primary
particles in the gap is usually different from the escape energy,
and more importantly, both $\lambda_{\rm sp}$ and $\lambda_{\rm pp}$
depend on $\epsilon$, but with different dependences. Furthermore, 
the characteristic $\epsilon$ from the gap also depends on pulsar 
parameters (e.g., $P$, $B_p$), which will alter the slope of the 
deathline.

Now we derive the PSD line for VG accelerators.  This is physically a
radio emission deathline for the anti-parallel rotators (APRs) (Zhang
\& Harding 2000). First, we neglect photon splitting and derive the
gap parameters following the same procedure as Zhang, Harding \&
Muslimov (2000). The difference here is that we derive the photon mean
free path by adopting the near-threshold condition for pair
creation. Assuming dipolar configuration at the surface, the curvature
radiation (CR) controlled gap has a height $h({\rm CR-VG})=3.2
\times 10^3 ({\rm cm}) P^{4/7} {B'}^{-3/7} \xi^{-2/7}$, and the
resonant inverse Compton scattering (ICS) controlled gap has a height
of $h({\rm ICS-VG})=3.3\times 10^3 ({\rm cm}) P^{3/7} {B'}^{-1}
\xi^{-4/7}$. Notice that although both gaps have a similar height for 
the typical values, the ICS gap height has a much steeper inverse
dependence on $B'$, which means that in slightly higher fields, ICS
becomes the dominant mechanism for pair production.
We therefore adopt ICS-VG parameters to define the photon splitting
deathline. The characteristic photon energy from such an accelerator,
which is the lowest photon energy produced within the gap by resonant
inverse Compton scattering of the electrons off the thermal photons
that can produce pairs if photon splitting is neglected, can be
expressed as
\be
\epsilon_c({\rm ICS-VG}) \simeq 5.5\times 10^4 P^{1/14} {B'}\xi^{-3/7}.
\label{epsilonc}
\ee
Plug Eq.(\ref{epsilonc}) into Eqs. (\ref{lambdasp}) and
(\ref{lambdapp}), and using
\be
\lambda_{sp} \leq \lambda_{pp}
\label{PSD}
\ee
as the definition of photon splitting dominating pair production, the
photon splitting deathline for the APRs is (see Fig.1)
\be
B_p \geq 1.0\times 10^{14} {\rm G} (P/ 1 {\rm s})
^{-10/49}\xi^{4/49}, 
\label{1}
\ee
or by adopting $B_p=6.4\times 10^{19} {\rm G} \sqrt{P\dot P}$,
$\dot P\geq 2.44\times 10^{-12} (P/ 1 {\rm s})
^{-69/49} \xi^{8/49}$.
This line has a steeper slope than the one derived using the 
``escape energy'' criterion (Baring \& Harding 2001), and allows 
two HBPs with relatively short periods, i.e., PSR J1119-6127 and PSR
J1726-3530, to be below the deathline, which means that they should
not be dead even if their gaps are (but not necessary are) vacuum type.
The reasons are that the faster pulsars tend to
have larger acceleration potentials and hence, produce more energetic
photons than the slower pulsars, and that for higher energy
photons, even higher magnetic field strength is required for photon
splitting to suppress pair production due to the different
$\epsilon$-dependences of both attenuation lengths (see
Eqs.[\ref{lambdasp}] and [\ref{lambdapp}]). Another remark is that
the deathline has a very weak dependence on $\xi$, thus the deathline
is a very nice indicator regardless of the location of the sparks in the
polar caps. We note that all known ``magnetars'' are above this
deathline, so that their radio quiescence is understandable if their
surface charges are bound to the surface so that the type of their
inner gaps is expected to be vacuum-like if they exist.

\section{Space-charge-limited flow models: are the gaps lengthened?}

PSR J1814-1744 is above the deathline of the APRs, and thus must be a
parallel rotator (PR) with a SCLF accelerator, if all three splitting 
modes operate (Zhang \& Harding 2000). In such a case, we can also
define a PSD line. But this line is no longer a ``deathline'', but
rather a line which determines whether the accelerator should be
lengthened with respect to its normal height (that derived assuming 
that photon splitting plays no role). SCLF accelerators with
frame-dragging effect taken into account have been explicitly studied
by Muslimov \& Tsygan (1992) and Harding \& Muslimov (1998). Despite
the complicated form of the parallel electric field in the gap, some
approximations could be made in certain regimes. More specifically,
for the solution with upper boundary condition, $E_\|$ increases
linearly with height in the beginning and gets saturated at a certain
level when the gap height is comparable to the polar cap radius
(Harding \& Muslimov 1998). For young pulsars, the $E_\|$ of the gaps
have not reached the saturated values, and the gaps are pancake-shaped. 
This is defined as Regime I. For older pulsars, $E_\|$ have attained 
the saturated values well below the pair formation front. The gap shape
in this case is long and narrow, and we define it as the Regime II 
SCLF gap. For both CR or ICS controlled
SCLF, we can have both the cases for regime I and regime II. We then
have four types of accelerators. Using the near-threshold pair
production condition which is applicable for HBPs, we have derived the
gap parameters of all these types. We find that in the phase space
where the three HBPs reside, the gaps are marginally in Regime II if
the gap is controlled by CR, but are in regime I if the gap is
controlled by ICS (PSR J1814-1744 is marginally in regime I). The
CR-controlled gap  has a height of 
$h({\rm CR-SCLF})=2.4\times 10^5{\rm cm} {B'}^{-3/4}P^{7/4}R_6^{-5/4}
\xi^{-1/2} (\cos\chi)^{-3/4}$, while the ICS-controlled gap has a
much lower height of $h({\rm ICS-SCLF})=8.1\times 10^3{\rm cm}
{B'}^{-14/15} P^{13/30} R_6^{3/10}\xi^{-8/15} (\cos\chi)^{-1/15}$ and
a lower potential, where $\chi$ is the pulsar obliquity. Therefore ICS
is the dominant pair production mechanism for the HBPs. Another
argument is that in a strong field a stable
ICS-controlled accelerator could be formed right above the surface
(Harding \& Muslimov 1998). 
The characteristic ICS photon energy produced from such a gap is
\be
\epsilon_c({\rm ICS-SCLF}) \simeq 2.3\times 10^4 {B'}^{14/15}P^{1/15}
R_6^{1/5}\xi^{-7/15} (\cos\chi)^{1/15},
\label{epsilonc2}
\ee
which is smaller than $\epsilon_c({\rm ICS-VG})$
(Eq.[\ref{epsilonc}]). The PSD line is therefore expected to be
lower. Notice that the dependence on $\cos\chi$ is rather weak.
Using a same criterion (Eq.[\ref{PSD}]), the PSD line for the
SCLF gaps is then (see Fig.1)
\be
B \geq 3.6\times 10^{13} {\rm G} (P/1{\rm s})^{-22/113} \xi^{4/113},
\label{2}
\ee
or $\dot P\geq 3.16\times 10^{-13} (P/1{\rm s})^{-157/113} \xi^{8/113}$.
Notice that this line is only a rough indication due to the analytic
approximation adopted here, and detailed numerical calculations are 
necessary. Nevertheless, with Eq.[\ref{2}] we find that all
the three known HBPs as well as some more pulsars are above this
line (Fig.1), which means that the delayed pair production conjectured
by Zhang \& Harding (2000) is necessary to interpret these pulsars. 
The gaps in these pulsars must be lengthened to allow pair formation
front to occur at a higher altitude. The deathline
for the HBPs (which should be PRs), is still the ``SCLF deathline''
defined according to the binding condition of the electrons, as
discussed in Zhang \& Harding (2000), which occurs around $B_p \simg
2\times 10^{14}$ G.

\begin{figure}
\epsfysize=8.5cm 
\hspace{2.5cm}\epsfbox{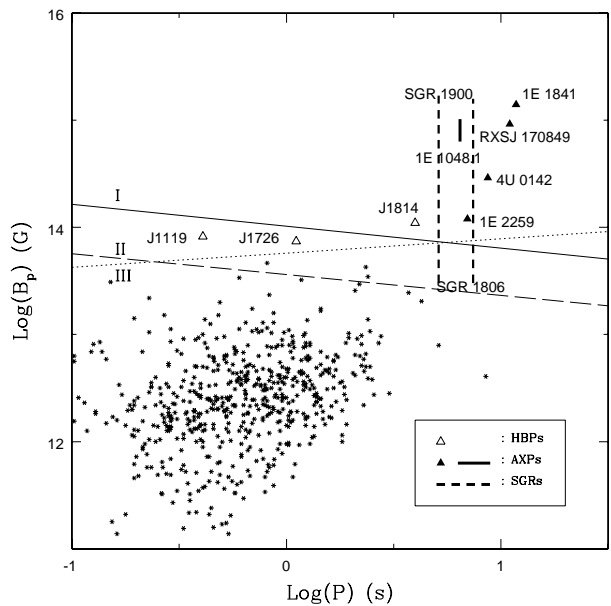} 
\caption[h]{A $B_p-P$ diagram of some pulsars and magnetars, and
different PSD lines. Line I is the PSD line for VGs, which is the
deathline of the APRs; Line II is the PSD line for SCLFs, which is the
critical line to define whether delayed pair formation is necessary;
Line III is the Baring \& Harding line, which is defined using the
escape energy criterion.}
\end{figure}

\section{Possible implications: connection with the AXPs/SGRs}

All the results presented here are based on the assumption that all
three photon splitting modes operate. Therefore the consequences from
this picture can be then regarded as the criteria to test the assumption
itself. All known HBPs as well as some  pulsars with lower fields do
need to invoke ``delayed pair production'' if their gaps are SCLF
types, as expected before (Zhang \& Harding 2000). The 
particle luminosity from such an accelerator therefore should be enhanced
with respect to that of a normal pulsar. This allows a direct distinction
between the two splitting scenarios, since different scenarios may
result in different $\gamma$-ray and X-ray luminosity
predictions. Further work is necessary to reveal such differences.
A caution is that the analytic approach adopted here may not be good
enough to describe the phase space where PSR J1814-1744 is located, so
that the ``photon splitting dominant line'' may be altered after
numerical calculations. Nonetheless, the qualitative conclusion that
``delayed pair production'' is necessary will not be changed.

If it turns out that in a magnetar environment, still only one
splitting mode can operate, some other ideas must be introduced to
account for the radio quiescence of the AXPs and the SGRs. One such
idea is that the particle flows induced by bursting activities may be
able to short out the inner accelerator of a magnetar for a long
period of time (Thompson 2000). In this picture, all AXPs must be
assumed to have experienced some bursting behaviors recently. 
Alternatively, radio quiescence in AXPs/SGRs is natural if these
objects are powered by accretion from their fossil disks.

\acknowledgements
We thank Matthew Baring for helpful discussions. B.Z. acknowledges
NASA NAG5-9192 and NAG5-9193 for support.



\end{document}